\documentclass[%
twocolumn,
superscriptaddress,
 amsmath,amssymb,
aps,
pra,
]{revtex4-2}

\usepackage{graphicx}
\usepackage{dcolumn}
\usepackage{bm}
\usepackage{hyperref}
\usepackage{color} 
\usepackage{txfonts}
\usepackage[dvipsnames]{xcolor}
\usepackage{ragged2e}
\usepackage{amsmath}

\newcommand{\ketbra}[2]{|#1\rangle\langle#1|}

\begin{document}

\title{Quantum heat engine based on 
a spin-orbit and Zeeman-coupled Bose-Einstein condensate}

\author{Jing Li}
\email{Corresponding author: jli@ucc.ie}
 \affiliation{Department of Physics, University College Cork, T12 H6T1 Cork, Ireland}

\author{E. Ya Sherman}%
\affiliation{Departamento de Química-Física, UPV/EHU, Apartado 644, 48080 Bilbao, Spain}%
\affiliation{IKERBASQUE, Basque Foundation for Science, 48011 Bilbao, Spain}
\affiliation{EHU Quantum Center, University of the Basque Country UPV/EHU}

\author{Andreas Ruschhaupt}
\affiliation{Department of Physics, University College Cork, T12 H6T1 Cork, Ireland}%

\begin{abstract}
We explore the potential of a spin-orbit coupled Bose-Einstein condensate for thermodynamic cycles. 
For this purpose we propose a quantum heat engine based on a condensate with spin-orbit and Zeeman coupling as a working medium.
The cooling and heating are simulated by contacts of the condensate with an external magnetized media and demagnetized media. We examine the condensate ground state energy and its dependence on the strength of the synthetic spin-orbit and Zeeman couplings and interatomic interaction. Then we study the efficiency of the proposed engine.
The cycle has a critical value of spin-orbit coupling  
related to the engine maximum efficiency.
\end{abstract}

\maketitle

\textit{Introduction} Quantum cycles are of much importance both for fundamental research and for applications in quantum-based technologies\cite{Goold_2016,Stevebook}. 
Quantum heat engines have been demonstrated in recent on several quantum platforms, 
such as trapped ions \cite{Lutz2012,Kilian2016}, quantum dots \cite{Sothmann_2014} and optomechanical oscillators \cite{Pierre2014,Markus2014,Elouard_2015,Gaub2002}.
Well-developed techniques for experimental control make Bose-Einstein condensates (BECs) \cite{pethick_smith_2008} 
a suitable system for a quantum working medium of a thermal machine \cite{Charalambous_2019,Myers_2022,Li2018}.

Recently, a quantum Otto cycle was experimentally realized using a
large quasi-spin system with individual cesium (Cs) atoms
immersed in a quantum heat bath made of ultracold rubidium (Rb) atoms \cite{widera2019,Widera21}. 
Several spin heat engines have been theoretically and experimentally implemented using a single-spin qubit \cite{Franco2021}, ultracold atoms \cite{Georges2013},
single molecule \cite{Berakdar2014}, 
a nuclear magnetic resonance setup \cite{Serra2019} 
and a single-electron spin coupled to a harmonic oscillator flywheel \cite{Poschinger2019}. 
These examples have motivated our exploration of the spin-orbit coupled BEC considered in this paper.

Spin-orbit coupling (SOC) links a particle’s spin to its motion, and artificially introduces charge-like physics into bosonic neutral atoms \cite{Zhang2016}. 
The experimental generation \cite{Spielman11,Hui2010,Shizhong2011,PhysRevA.84.025602} of SOC is usually accompanied by a Zeeman field, which breaks various symmetries of the underlying system and induces interesting quantum phenomena, e.g. topological transport\cite{Niu2010}. 
In addition, in the spin-orbit coupled BEC system, more studies on moving solitons \cite{Pelinovsky13,Wu13,Vladimir2017}, vortices \cite{PhysRevA.84.063604}, stripe phase \cite{PhysRevLett.107.270401} and dipole oscillations \cite{Pan2012} have been reported.

In this paper, we propose a BEC with SOC as a working medium in a quantum Stirling cycle.
The classic Stirling cycle is made of two isothermal branches, connected by two isochore branches.
The BEC is characterized by SOC, Zeeman splitting, a self-interaction, and is located in a quasi-one-dimensional vessel with a moving piston
that changes the length of the vessel.  
The external "cooling" and "heating" reservoirs are modelled by the interaction of the spin-1/2 BEC with an 
external magnetized and demagnetized medias. The expansion and compression works depend on the SOC and Zeeman coupling.
A main goal is to examine the condensate ground state energy and its dependence on the strength of the synthetic spin-orbit, Zeeman couplings,  interatomic interaction and length of the vessel.
For the semiquantitative analysis, perturbation theory is applied to understand the effects of SOC and Zeeman splitting.
We further analyze several important parameters and investigate how they affect the efficiency of the cycle, e.g. the critical
SOC strength for different self-interactions.

\textit{Model of the heat engine: Working medium}
We consider a quasi-one dimensional BEC, extended along the $x-$axis and tightly confined in the orthogonal directions.
 The mean-field energy functional of the system is then given by $E = \int_{-\infty}^{+\infty}\varepsilon dx$ with spin-independent self-interaction  of the Manakov's symmetry \cite{Manakov1974}:
\begin{eqnarray}
\varepsilon\!=\! \Psi^{\dagger} \mathcal{H}_0 \Psi + \frac{g}{2}(|\psi_\uparrow|^2+|\psi_\downarrow|^2)^2,
\label{hamfull}
\end{eqnarray}
where $\Psi \equiv (\psi_\uparrow,\; \psi_\downarrow)^{\mathrm T}$ (here ${\mathrm T}$ stands for transposition) and the wavefunctions $\psi_\uparrow$ and $\psi_\downarrow$ are related to the two pseudo-spin components. The parameter
$g$ represents the strength of the atomic interaction which can be tuned by atomic $s-$wave scattering 
length using Feshbach resonance \cite{PhysRevA.46.R1167,Feshbach98} with $g>0$, $g<0$, and $g=0$ giving the repulsive, attractive, and no atomic interaction, respectively. 
The Hamiltonian $\mathcal{H}_0$ in Eq. \eqref{hamfull} of the spin-1/2 BEC, trapped in an external potential $V(x)$, is given by
\begin{eqnarray}
\mathcal{H}_0={\frac{\hat{p}^{2}}{2m} \hat{\sigma}_{0} + \frac{\alpha}{\hbar}\hat{p}\hat{\sigma}_{x}}+\frac{\hbar}{2}\Delta\hat{\sigma}_z + V(x),
\label{ham0}
\end{eqnarray}
with $\hat{p}=-i\hbar\partial_{x}$ being the momentum operator in the longitudinal direction, $\hat{\sigma}_{x,z}$ being the
Pauli matrices, and $\hat{\sigma}_{0}$ being the identity matrix.
Here $\alpha$ is the SOC constant and $\Delta$ is the Zeeman field. 
We choose a convenient length unit $\xi$, an energy unit $\hbar^2/(m \xi^{2})$ and a time unit $m \xi^2/\hbar$ {and express the following equations in the corresponding dimensionless variables.}
The coupled Gross-Pitaevskii equations are now given by
\begin{eqnarray}\label{coupled_GPE}
i\frac{\partial }{\partial t}\psi_\uparrow &=&\left( -\frac{1}{2}\frac{\partial ^{2}}{%
\partial x^{2}}+\frac{\Delta }{2}+g \;n(x) +V(x)%
\right) \psi_\uparrow-i\alpha \frac{\partial }{\partial x}\psi_\downarrow,
 \\
i\frac{\partial }{\partial t}\psi_\downarrow &=&\left( -\frac{1}{2}\frac{\partial ^{2}}{%
\partial x^{2}}-\frac{\Delta }{2}+g \; n(x)+V(x)%
\right) \psi_\downarrow-i\alpha \frac{\partial }{\partial x}\psi_\uparrow,
\end{eqnarray}
where the density is given by $n(x)=|\psi_\uparrow|^2+|\psi_\downarrow|^2$. We fix the norm $N=\int_{-\infty}^{\infty} n(x)\, dx  = 1$.

We consider a hard-wall potential $V(x)$ of half width $a$: 
\begin{equation}\label{potential}
V(x)  = 0, \quad  (|x| \leq a), \qquad  V(x)= \infty \quad (|x| > a).
\end{equation}
This potential is analogous to a piston in a thermodynamic cycle and it allows one to define the work of the quantum cycle.
The ground state $\Psi$ of the BEC then depends on the half width $a$, the detuning $\Delta$, the interactions $g$ and the SOC $\alpha$, i.e. $\Psi_{\alpha,g} (a, \Delta)$, and 
the corresponding total ground state energy of the BEC is then denoted as $E_{\alpha,g}(a, \Delta)$.
We define also the pressure $P_{\alpha,g}(a, \Delta)$ as a measure 
of the energy $E_{\alpha,g}(a, \Delta)$ stored per total length $2a$:
\begin{equation}\label{pressure}
   {P_{\alpha,g}(a, \Delta) \equiv -\frac{\partial E_{\alpha,g}(a,\Delta)}{2\partial a}.}
\end{equation}

In the special case of $\Delta=0$
and for the spin-independent 
self-interaction proportional to $n(x)$, the energy \cite{PhysRevB.74.153313,Tokatly2010} 
is given by $E_{\alpha,g}(a,0) = E_{0,g}(a,0)-\alpha^{2}/2$ resulting in
$\alpha-$independent pressure $P_{\alpha,g}(a,0)$. 
Notice that at both nonzero $\alpha$ and $\Delta$, the system is characterized by a magnetostriction in the form
${\cal M}_{\alpha,g}(a,\Delta)=\partial P_{\alpha,g}(a,\Delta)/\partial \Delta.$

\begin{figure}[t]
\begin{center}
\includegraphics[width=0.8\columnwidth]{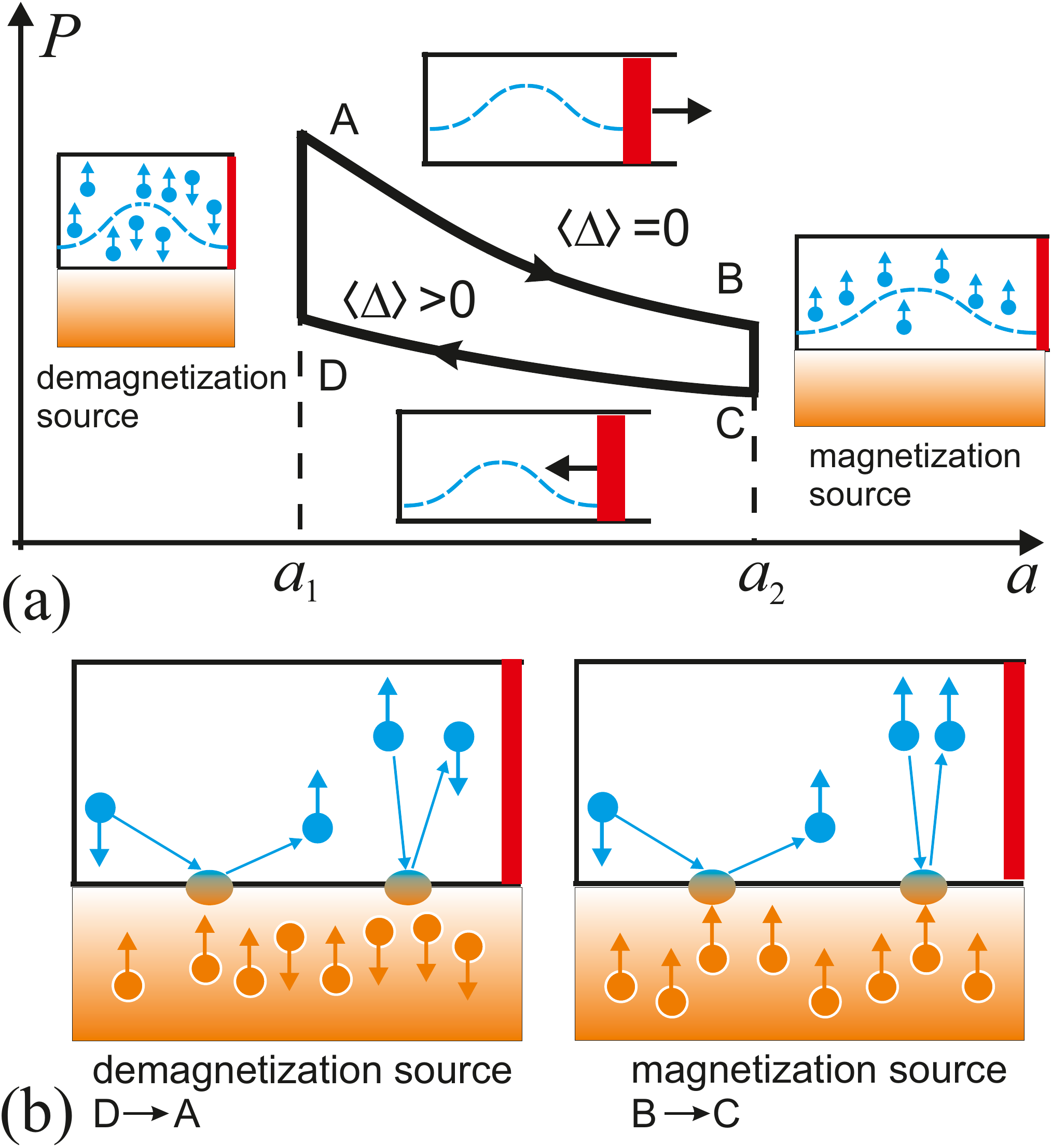}
\end{center}
\vspace*{-2mm}
\caption{ (a) The schematic diagram of the quantum Stirling cycle based on the Zeeman and SOC. (b) Visualization of the demagnetization (left) and magnetization (right) processes with the external sources; the blue dots represent the BEC atoms and the orange dots represent the external source.
\label{fig1:cycle}}
\end{figure}

%
\textit{Model of the heat engine: Quantum Stirling cycle} 
We consider a quantum Stirling cycle keeping 
the interaction $g$ and the SOC $\alpha$ fixed during the whole process. The key idea is that 
the external "cooling" and "heating" reservoirs are modelled by the interaction of the spin 1/2 BEC with an 
external magnetized media (see Fig. \ref{fig1:cycle}(b), right) resp. demagnetized media (see Fig. \ref{fig1:cycle}(b), left).
This external, (de)magnetized source leads to a random magnetic field in the condensate and because of the Zeeman-effect this corresponds to a detuning of the condensate to $\Delta$ with some probability density distribution $p(\Delta)$.
We assume that this external source brings the system to a stationary state 
with the condensate described by a density operator
\begin{eqnarray}
\label{rho}
\hat{\rho} = \int p(\Delta) \,
\ketbra{\Psi_{\alpha,g} (a, \Delta)}{\Psi_{\alpha,g} (a, \Delta)}\, d\Delta .
\end{eqnarray}
The probability density distribution of the demagnetizing source $p_{\rm{dm}} (\Delta)$ is centered around $ \langle \Delta \rangle_{\rm dm} \equiv \int \, \Delta \, p_{\rm dm}(\Delta)\,d\Delta =0$ while the one of the magnetizing source $p_{\rm{m}} (\Delta)$ is centered around a positive value $\langle \Delta \rangle_{\rm m}>0$. As an increase in $\Delta$ decreases the BEC energy \cite{pethick_smith_2008} by an $\alpha-$dependent amount, the demagnetization source plays the role of a ``hot thermal bath'' here and the magnetization source plays the role of a ``cold thermal bath". In general there could exist a stationary external magnetic field leading to an additional detuning during the cycle.  We neglect this possibility in the following in order to simplify the notation.

The realization of the Stirling cycle is described by a four-stroke protocol, illustrated in Fig. \ref{fig1:cycle}(a).
We start at point $A$ with the BEC being in contact with the demagnetization source,
leading to an effective detuning centered around $\langle \Delta \rangle_{\rm dm}=0$.
The potential is of half width $a_1$. The BEC state is given by Eq. \ref{rho} with $p(\Delta) = p_A (\Delta) \equiv p_{\rm dm}(\Delta)$.

\textit{Quantum ``isothermal" expansion stroke $A\rightarrow\,B$:} During this stroke, the working medium stays in contact with the external demagnetization source while the potential expands adiabatically from $a_{1}$ to $a_{2}$ without excitation in the BEC. The probability density distribution $p(\Delta)$ stays constant during this "isothermal" stroke, i.e. we have $p_{A}(\Delta)=p_{B}(\Delta) = p_{\rm{dm}} (\Delta)$ (effective detuning centered around $\langle \Delta \rangle_{\rm dm}=0$).
The average work done during this ``isothermal" expansion stroke can be then calculated as \cite{Hanggi2003}
$\langle W_{e} \rangle = \int  \, p_{\rm{dm}} (\Delta)
\left(E_{\alpha,g} (a_1, \Delta) - E_{\alpha,g} (a_2, \Delta)\right)d\Delta$.

\textit{Quantum isochore cooling stroke $B\rightarrow C$:}
The contact with the demagnetization source is switched off and the working medium is brought into contact with the magnetization source while keeping $a_2$ constant.
The probability distribution $p(\Delta)$ is changed to $p_C(\Delta)\equiv p_{\rm m}(\Delta)$, this corresponds to a "cooling" (as the total energy of the BEC is lowered).
The average heat exchange in this stroke can be calculated as
$\langle Q_{c} \rangle = \int\, \left(p_{\rm m} (\Delta) - p_{\rm dm} (\Delta) \right) E_{\alpha,g} (\Delta, a_2) d\Delta$.

\textit{Quantum ``isothermal" compression stroke $C\rightarrow\,D$:}
During this stroke, the working medium stays in contact with the external magnetization source while the BEC compresses adiabatically from potential half width $a_{2}$ to $a_{1}$ without excitation in the BEC. The probability density distribution $p(\Delta)$ remains constant during this "isothermal" stroke, i.e. we have $p_D(\Delta)=p_C(\Delta) = p_{\rm m} (\Delta)$ leading to an effective detuning centered around $\langle \Delta \rangle_{\rm m}>0$. The average work done during this ``isothermal" compression is
$\langle W_{c} \rangle =  \int\, p_{\rm m} (\Delta)
\left(E_{\alpha,g} (a_2, \Delta) - E_{\alpha,g} (a_1, \Delta)\right) d\Delta$.

\textit{Quantum isochore heating stroke $D\rightarrow A$:}
The contact with the magntetization source is switched off and the working medium is brought again into contact with the demagnetization source while keeping $a_1$ constant. The probability distribution $p(\Delta)$ is changed back to $p_A(\Delta)=p_{\rm dm}(\Delta)$, this corresponds to a "heating" (as the total energy of the BEC is increased). The average heat exchange in this stroke can be calculated as
$\langle Q_{h} \rangle =  \int\, \left(p_{\rm dm} (\Delta) - p_{\rm m} (\Delta) \right) E (\Delta, a_1) d\Delta$.

To study this quantum cycle, {it is important to examine and understand} the dependence of the BEC ground-state energy on the different parameters. This will be done in the following.

\textit{Perturbation theory for the ground state energy} The complex BEC system used in the thermodynamic cycle does not have an exact analytical solution.
However, we can obtain analytical insight by considering perturbation theory of the ground state energy $E_{\alpha,0}(a,\Delta)$ of the non-selfinteracting BEC (i.e. $g=0$) at small $\alpha$ (and nonzero $\Delta$), as well as at small $\Delta$ (and nonzero $\alpha$). 

In the case of small $\alpha$ then $\alpha\ll\,1/a$, the Hamiltonian in Eq. \eqref{hamfull} can be written as $\mathcal{H}_{0} = \mathcal{H}_{0,0} + \mathcal{H}_0^{\prime}$   
where $\mathcal{H}_{0}=\hat{p}^{2}/2+\Delta\hat{\sigma}_{z}/2+V(x)$ and the perturbation term being $\mathcal{H}_0^{\prime}=\alpha \hat{p} \hat{\sigma}_{x}$.
The eigenstate basis of $\mathcal{H}_{0,0}$ is given by 
$\psi_{n,\downarrow}^{(0)}(x) = \left[0,\psi_{n}(x)\right]^{\rm T}$,
$\psi_{n,\uparrow}^{(0)}(x) = \left[\psi_{n}(x),0\right]^{\rm T}$,
where $\psi_{n}(x)$ are the eigenstates of the potential in Eq.\eqref{potential}. 
The first-order 
correction to the energy vanishes and the second-order correction becomes:
\begin{equation}
{\epsilon^{(0)}_{2}=
 -\sum_{n>1}
 \frac{\lvert\langle \psi_{n,\uparrow}^{(0)}(x)|\mathcal{H}^{\prime}_{0}|\psi_{0,\downarrow}^{(0)}(x)\rangle\rvert^2}
 {(n^{2}-1)\pi^{2}/(8a^{2})+\Delta}.}
\end{equation}
Thus, the total ground state energy $E_{\alpha,0} (a,\Delta)$ of the system up to second order in $\alpha$ is given by
\begin{equation}\label{alpha_pertu}
    E_{\alpha,0} (a,\Delta) \approx 
\frac{\pi^2}{8a^2}-\frac{\Delta}{2} - \frac{\pi^2\alpha^2}{4 \Delta a^2}  +
\frac{\pi^2\alpha^2}{8a^{4}\Delta^{2}}
\chi(a,\Delta)\cot\frac{\chi(a,\Delta)}{2},
\end{equation}
where $\chi(a,\Delta)\equiv\,\sqrt{\pi ^2-8 a^2\Delta}.$
We can simplify Eq. \eqref{alpha_pertu} by approximating the expression up to first order in $\Delta$:
\begin{equation}
\label{Ealpha}
    E_{\alpha,0} (a,\Delta) \approx \frac{\pi ^2}{8 a^2}-\frac{\Delta}{2}-\frac{\alpha ^2}{2}+
   {\frac{\pi^{2}-6}{3\pi^{2}} 
    \left(\frac{a}{\ell_{\rm sr}}\right)^{2}\Delta.}
\end{equation}
The first three terms on the right-hand side
of Eq. \eqref{Ealpha}
correspond to kinetic energy, Zeeman energy (at $\alpha=0$) and SOC energy (at $\Delta=0$). Here we introduced the spin rotation length $\ell_{\rm sr}\equiv\,1/\alpha$ with $a/\ell_{\rm sr}\ll\,1$.

Alternatively, in the case of large $\alpha$ then $\alpha>1/a$ and small detuning $\Delta$,
the Hamiltonian can be written as $\mathcal{H}_{0} = \mathcal{H}_{0,1} + \mathcal{H}_1^{\prime}$   
where $\mathcal{H}_{0}=\hat{p}^{2}/2+\alpha \hat{p} \hat{\sigma}_{x}+V(x),$ and the perturbation term
$\mathcal{H}_1^{\prime}=\Delta\;\hat{\sigma}_{z}/2.$ 
The unperturbed $\mathcal{H}_{0,1}$ has pairs of degenerate eigenstates $\psi_{a}^{(0)}$ and $\psi_{b}^{(0)}$ with the energy $E_{\alpha,0}(a,0)$:
\begin{eqnarray}
 \psi_a^{(0)}(x) = \psi_{n}(x)e^{-\mathrm{i}\alpha x}\begin{pmatrix}
 1\\
1
\end{pmatrix},\;\;
 \psi_b^{(0)}(x) = \psi_{n}(x)e^{\mathrm{i}\alpha x}\begin{pmatrix}
 1\\
-1
\end{pmatrix}.
\end{eqnarray}
Based on the perturbation theory for degenerate states and taking into account that 
the diagonal matrix elements of the perturbation, $\Delta\langle \psi_{i}^{(0)} |\sigma_{z}| \psi_{i}^{(0)}\rangle/2=0$, 
we obtain at $a/\ell_{\rm sr} \gg\,1$ the ground state energy in the form:
\begin{equation}
\label{delta_pertu}
  E_{\alpha,0}(a,\Delta) \approx
  \frac{\pi^2}{8a^{2}}-\frac{\alpha^2}{2} - \frac{\pi^{2}}{4}
  \frac{\ell_{\rm sr}/a}{\left|\left(2 a/\ell_{\rm sr}\right)^{2}-\pi^{2}\right|}
    \left|\sin \left(\frac{2a}{\ell_{\rm sr}}\right)\right|\Delta.
\end{equation}
When we look at the corresponding pressure following from Eq. \eqref{delta_pertu}, 
we can calculate approximately
the pressure difference $\delta P$ between the points $B$ and $C$ in the cycle ({at $a_2$,} see Fig. \ref{fig1:cycle}). 
The difference $\delta P$ jumps from negative to positive
at certain widths where $2 a_{2}/ \ell_{\rm sr} \approx (n+1) \pi$ or $\alpha \approx (n+1) \pi /(2a_2)$ with $n=1,2,3,4,...$.
In addition, there is always an $\alpha$ between two consecutive ``jump points" where $\delta P$ becomes zero.
We will denote the first corresponding value of $\alpha,$
where the change of $\delta P$ for negative to positive occurs, as the critical $\alpha_c(g,\Delta)$.

\begin{figure}[t]
\begin{center}
\includegraphics[width=0.85\columnwidth]{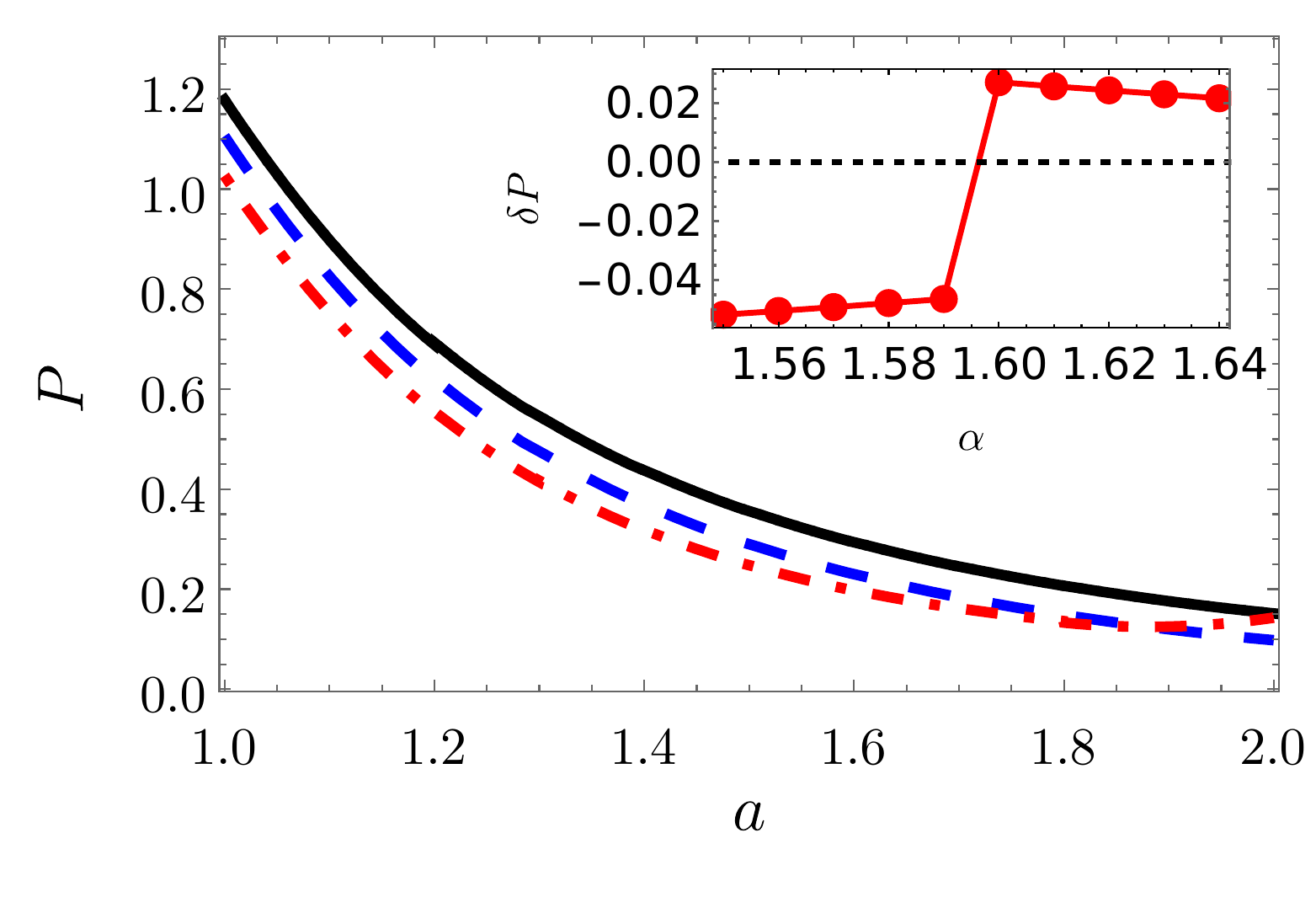}
\end{center}
\vspace*{-7mm}
\caption{Pressure $P_{\alpha,0}(a_{2},\Delta)$ versus potential half width $a$ for the cases of $\Delta =0,\alpha=1.6$ (solid black, essentially, $\alpha$-independent), $\Delta =1,\alpha=1$ (dashed blue) and $\Delta =1,\alpha=1.6$ (dot-dashed red).  (Inset) The pressure difference between points $C$ and $B$, $\delta P=P_{\alpha,0}(a_{2},1)-P_{\alpha,0}(a_{2},0)$. 
\label{fig2:PV}}
\end{figure}

\begin{figure}[t]
\includegraphics[width=0.85\columnwidth]{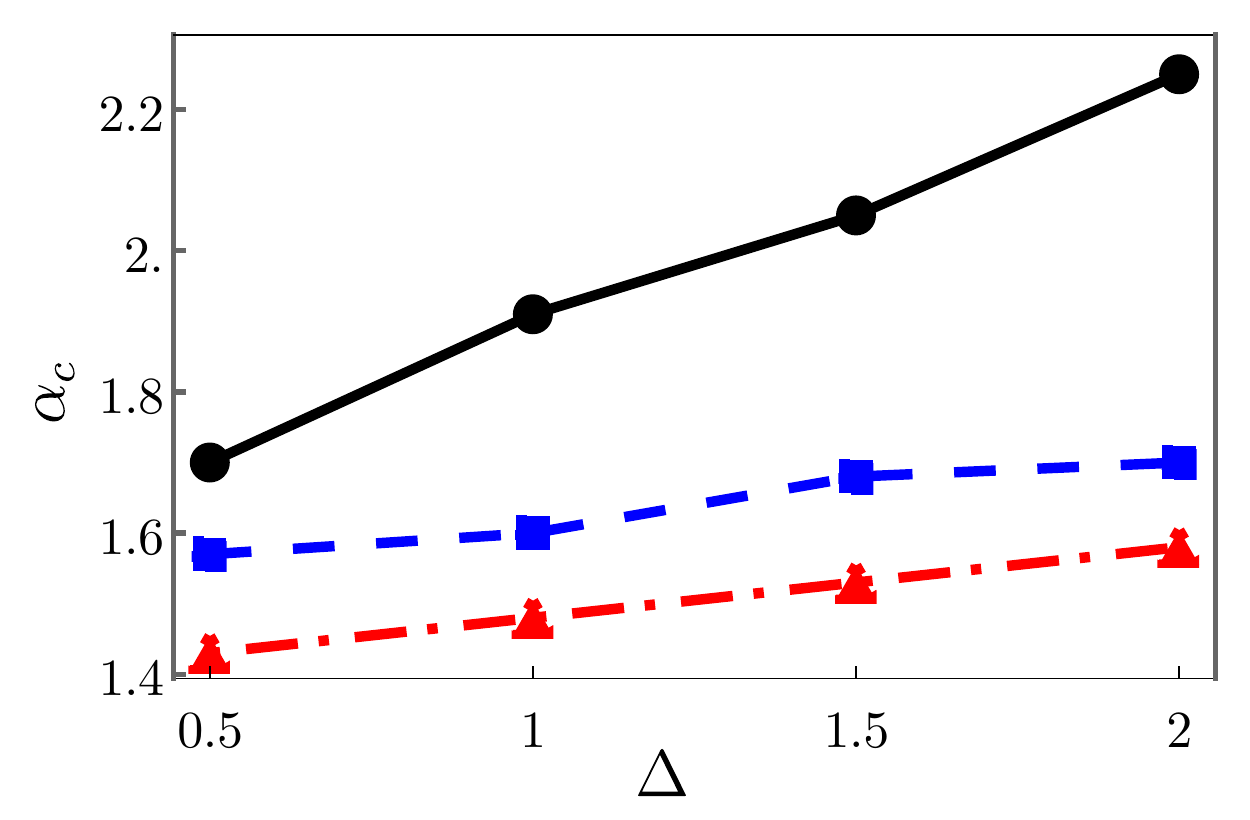}
\begin{picture}(0,0)
{\line(0,0){0.8}}
\end{picture}
\caption{Critical $\alpha_c(g,\Delta)$ versus detuning $\Delta$ for different nonlinearities: 
attractive $g=-1$ (black solid line), non-interaction $g=0$ (blue dashed line), 
and repulsive $g=1$ (dot-dashed red line).
\label{fig3:alphac_delta}}
\end{figure}
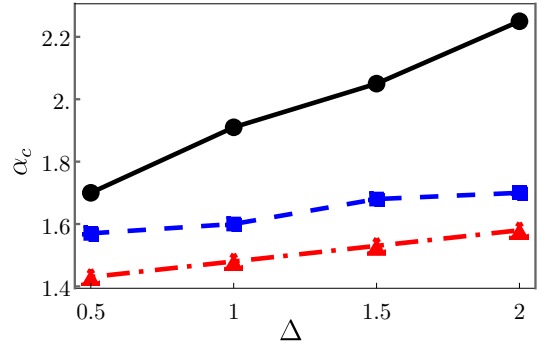

\textit{Energy and pressure }
{We examine now the exact numerical values of energy and pressure where 
we fix $a_1=1$ and $a_2=2$.
The corresponding pressure is} illustrated in Fig. \ref{fig2:PV} for a non-interacting BEC ($g=0$).
The shown pressure $P_{\alpha,0}(a_{2},0)$ for $\Delta=0$ does not depend on the strength of SOC $\alpha$ {as discussed above}. We can also see that the pressure $P_{\alpha,0}(a_{2},1)$
is {approximately} equal to the pressure $P_{\alpha,0} (a_{2},0)$ at $\alpha\,a_{2}\approx \pi,$ providing crossing of the red  and dotted lines; this corresponds then to a critical $\alpha_c(0,1)\approx 1.6$.
The corresponding difference in pressure $\delta P$ is shown in detail in the inset; it can be seen that $\delta P$ changes from negative to positive at
$\alpha_c(0,1)$ as one expects it from the perturbation theory above.

In Fig. \ref{fig3:alphac_delta}, the relations between the critical $\alpha_c(g,\Delta)$ and detuning $\Delta$
for different nonlinearities $g$ are plotted. From the perturbation theory for $g=0$ and for small $\Delta$, one expects a value of $\alpha_{c}(g,\Delta) \approx \pi/a_2 \approx 1.57$. 
The figure shows that {the exact} $\alpha_c$ is increasing with increasing $\Delta$ for all cases of $g$. 
There is a competition between SOC and Zeeman field, therefore, a larger detuning $\Delta$ requires automatically a larger $\alpha$ (and therefore a larger $\alpha_c$) to have an effect. We also see that $\alpha_c$ is larger (smaller) for attractive $g=-1$ (repulsive $g=1$) for all $\Delta$.
{The heuristic reason is that} there is a (kind of) compression (expansion) of the wavefunction for $g<0$ ($g>0$) and, therefore, a weaker (stronger) effect of SOC.
This requires heuristically a larger (smaller) $\alpha$ (and therefore $\alpha_c$) to show an effect.

\textit{Work, heat and efficiency of the engine}
Here we are mainly interested in the properties of the cycle originating from the BEC and not {in the details of the (de)magnetization source.}
Therefore, we assume that the probability density distributions $p_{\rm dm}$ resp. $p_{\rm m}$ are strongly peaked around
$\langle \Delta \rangle_{\rm dm}=0$
resp. $\langle \Delta \rangle_{\rm m} = \Delta_0 > 0$ such that we approximate $p_{\rm dm} (\Delta) = \delta(\Delta)$ and $p_{\rm m} (\Delta) = \delta(\Delta - \Delta_0)$ (where $\delta$ is the Dirac distribution).
In this case, the black-solid line and the blue-dashed line in Fig. \ref{fig2:PV} present an example 
of the expansion and compression strokes of the cycle shown in the schematic Fig. \ref{fig1:cycle}. The work done during the ``isothermal" expansion process in Fig. \ref{fig1:cycle}, $\langle W_{e} \rangle$, is then given by the energy differences:
$ \langle W_{e} \rangle =E_{\alpha,g} (a_{1},0)-E_{\alpha,g} (a_{2},0)$.
The cooling heat exchange from $B$ to $C$ $\langle Q_{c} \rangle$ through contact with the magnetization source, becomes
$        \langle Q_{c} \rangle=E_{\alpha,g} (a_{2},\Delta_0)-E_{\alpha,g} (a_{2},0).$
The work $\langle W_{c} \rangle$ done during the compression stroke is then $\langle W_{c} \rangle = E_{\alpha,g} (a_{2},\Delta_0)-E_{\alpha,g} (a_{1},\Delta_0)$. The heat in the last stroke can be calculated by $\langle Q_h \rangle = E_{\alpha,g} (a_{1},0)-E_{\alpha,g} (a_{1},\Delta_0)$.
The total work then becomes:
\begin{equation}
{\mathcal A}=\langle W_{c} \rangle+\langle W_{e}\rangle=\oint_{\rm ABCD} {P_{\alpha,g}(a,\Delta_0)da.}    
\end{equation}
For small $\Delta_0$,
\begin{equation}
{\mathcal A}=-\Delta_0 \int_{2a_{1}}^{2a_{2}} 
{\cal M}_{\alpha,g}(a,\Delta_0 \rightarrow\,0)da    .
\end{equation}
As defined above, at $\alpha=\alpha_{c}(g,\Delta_0)$, the pressures at $a_{2}$ for $\Delta=0$ and $\Delta_0>0$ approximately coincide.
If $\alpha>\alpha_{c}(g,\Delta_0),$ the pressure-dependencies on $a$ for $\Delta=0$ and $\Delta_0>0$ cross at a certain half width $\widetilde{a}$ with $a_{1}<\widetilde{a}<a_{2}$. In that case, the work done at the interval $(\widetilde{a},a_{2})$ provides a negative contribution while the contribution 
of the interval $(a_{1},\widetilde{a})$ can still increase.
In the following, we restrict our analysis to the case $\alpha \le \alpha_{c}(g,\Delta_0)$ while we expect a maximum of the total work close to $\alpha_{c}(g,\Delta_0)$.

The efficiency of each quantum cycle is now defined as
\begin{eqnarray}
\label{eff}
\eta = \frac{\mathcal A}{\langle Q_{h}\rangle}.
\end{eqnarray}
At small $\alpha\ll 1/(2a_{2})$ we may approximate the efficiency of the quantum cycle in terms of $\Delta_0$ as
\begin{eqnarray}\label{eff_pert}
   \eta \approx \left[\frac{\pi ^2}{2\Delta_0 ^2}\left(\frac{1}{a_{1}^2}-\frac{1}{a_{2}^2}\right)   + \frac{\pi ^2 }{4 \Delta_0 ^3} \zeta \right] \alpha^2,
\end{eqnarray}
where the coefficient $\zeta$ is
\begin{equation}\label{zeta}
 \zeta = 
 \frac{\chi(a_{1},\Delta_0)}{a_{1}^4}\cot\frac{\chi(a_{1},\Delta_0)}{2} -\frac{\chi(a_{2},\Delta_0)}{a_{2}^4}\cot\frac{\chi(a_{2},\Delta_0)}{2}.
\end{equation}
In the limit of $\Delta_0 \rightarrow 0$, the efficiency $\eta$ simplifies to 
\begin{eqnarray}
\label{zeroDelta}
 \eta =  \frac{2 \left(\pi ^2-6\right) }{3 \pi ^2}\left(a_{2}^2-a_{1}^2\right)\alpha ^2.
\end{eqnarray}
It is worth noticing that Eq. \eqref{zeroDelta} has two limits
with respect to the value of $a_{2}.$
Let us define $\eta_c$ as the efficiency at the critical $\alpha_c$. First, Eq. \eqref{zeroDelta} is applicable only at $\alpha\,a_{2}<\pi,$ 
thus, limiting the critical $\eta_{c}$ to the values of the order of 0.1. 
Secondly, for $g<0$ the value of $a_{2}$
is limited to $2/|g|$ \cite{Sulem_1999}, thus
$\eta_{c}$ is limited correspondingly. (Note that Eq. \eqref{zeroDelta} is not directly applicable to $g\ne\,0$ BEC).

\begin{figure}[t]
\begin{center}
\includegraphics[width=0.85\columnwidth]{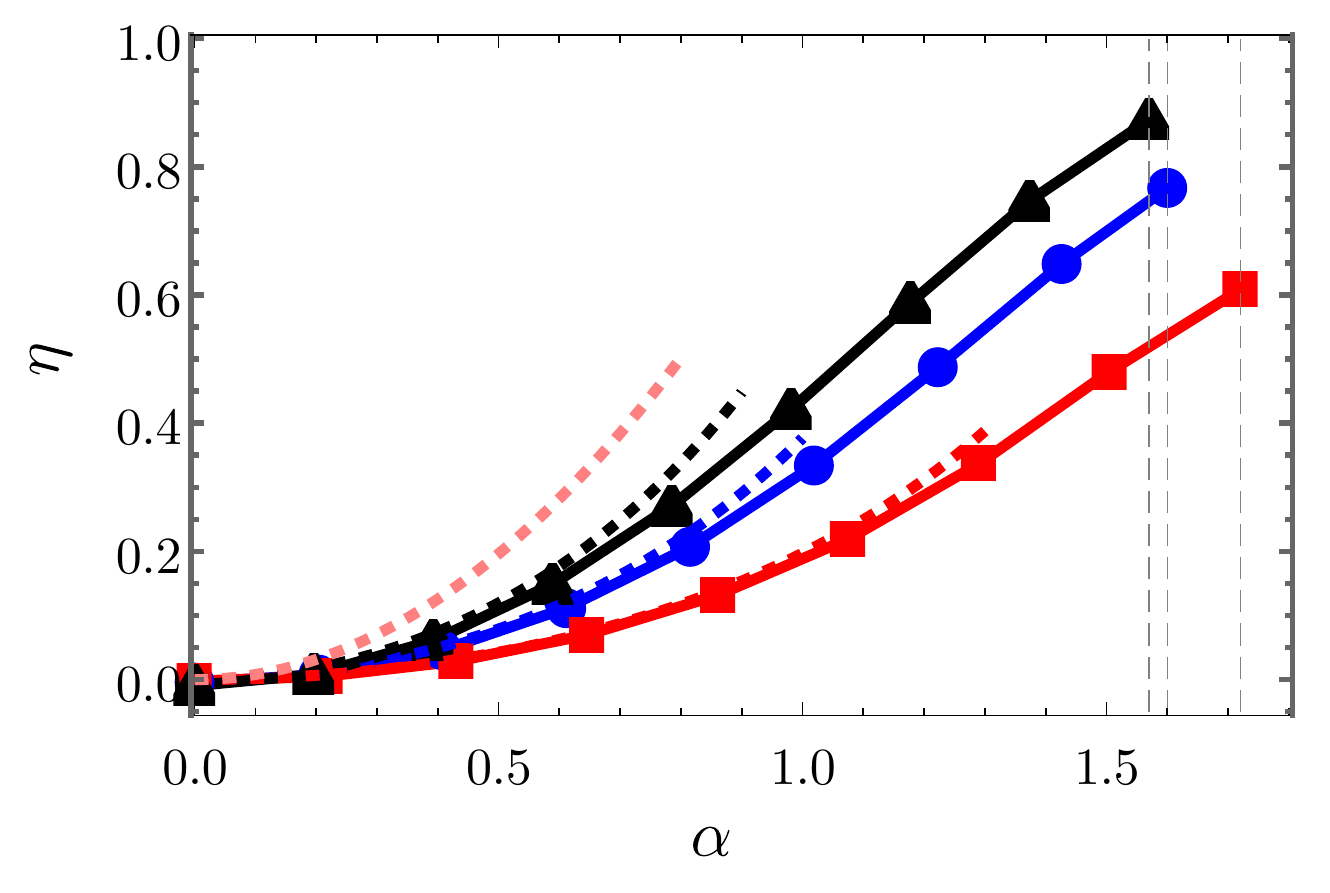}\\
\includegraphics[width=0.85\columnwidth]{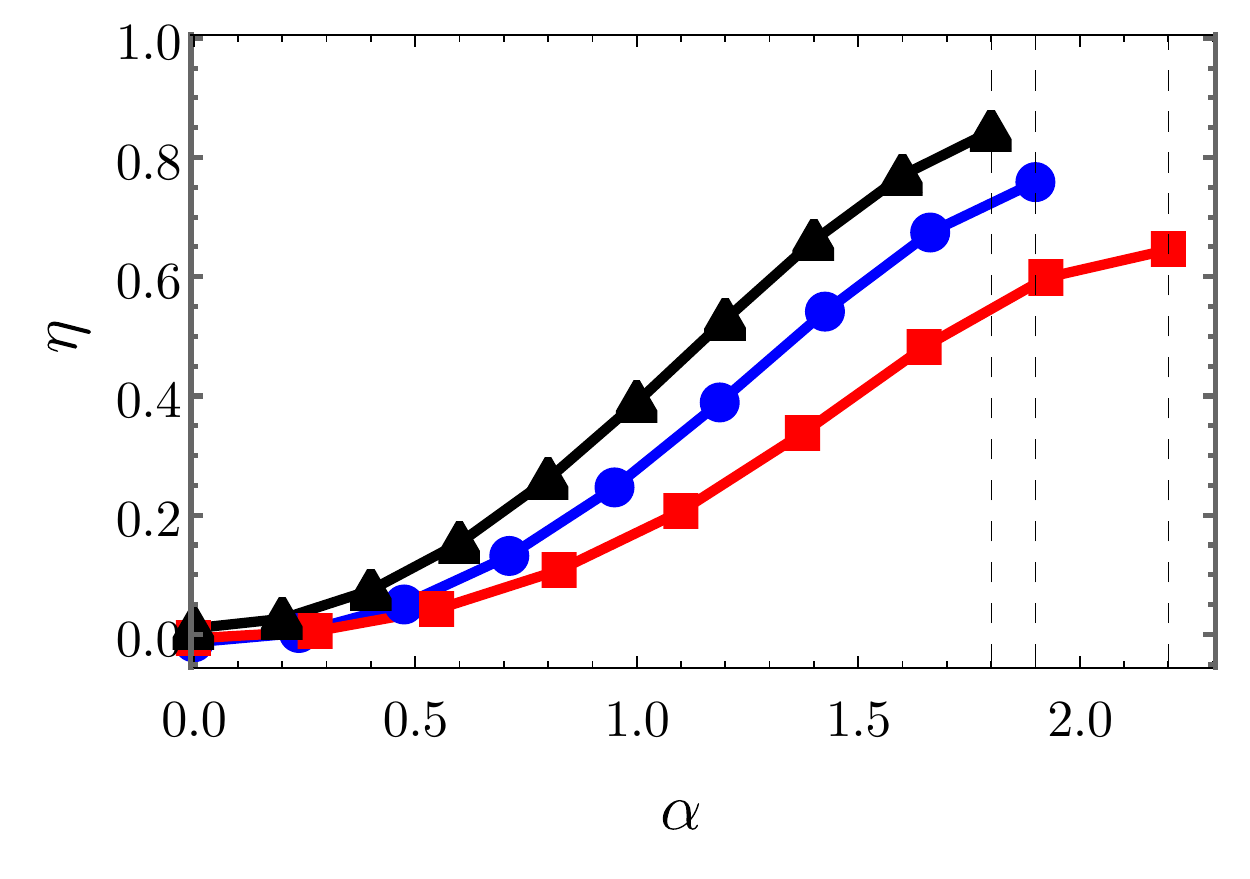}
\end{center}
\begin{picture}(0,0) 
\put(-70,285){\large (a)} 
\put(-68,143){\large (b)} 
\end{picture}
\vspace*{-10mm}
\caption{Efficiency $\eta$ versus $\alpha$ with $\Delta_0=0.5$ (solid black), $\Delta_0=1.0$ (solid blue) and $\Delta_0=2.0$ (solid red); the dotted vertical lines denote the critical SOC strength $\alpha_{c}(g,\Delta_{0})$.
(a) $g=0$; results based on perturbation theory in Eq. (\ref{eff_pert}) (blue, red and black dashed lines); the dashed pink line is given by Eq. (\ref{zeroDelta})  for $\Delta_0 \to 0$.
(b) $g=-1$. \label{fig4:effi_gs_pert}}
\end{figure}
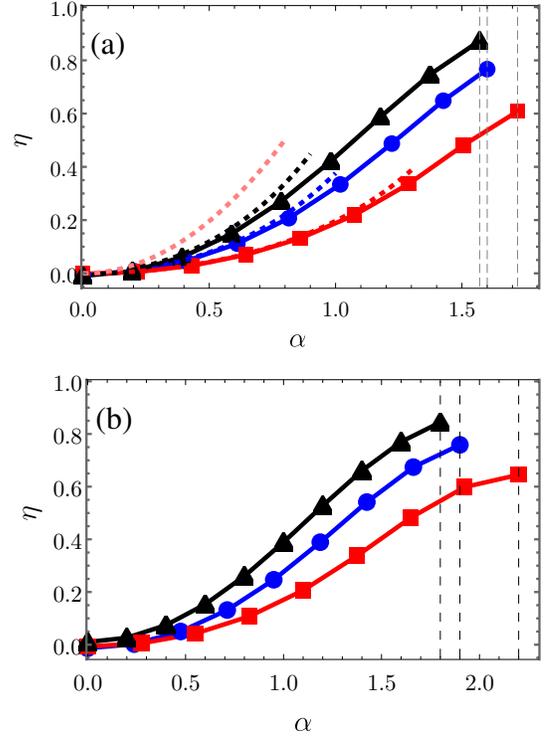

\begin{figure}[t]
\begin{center}
\includegraphics[width=0.85\columnwidth]{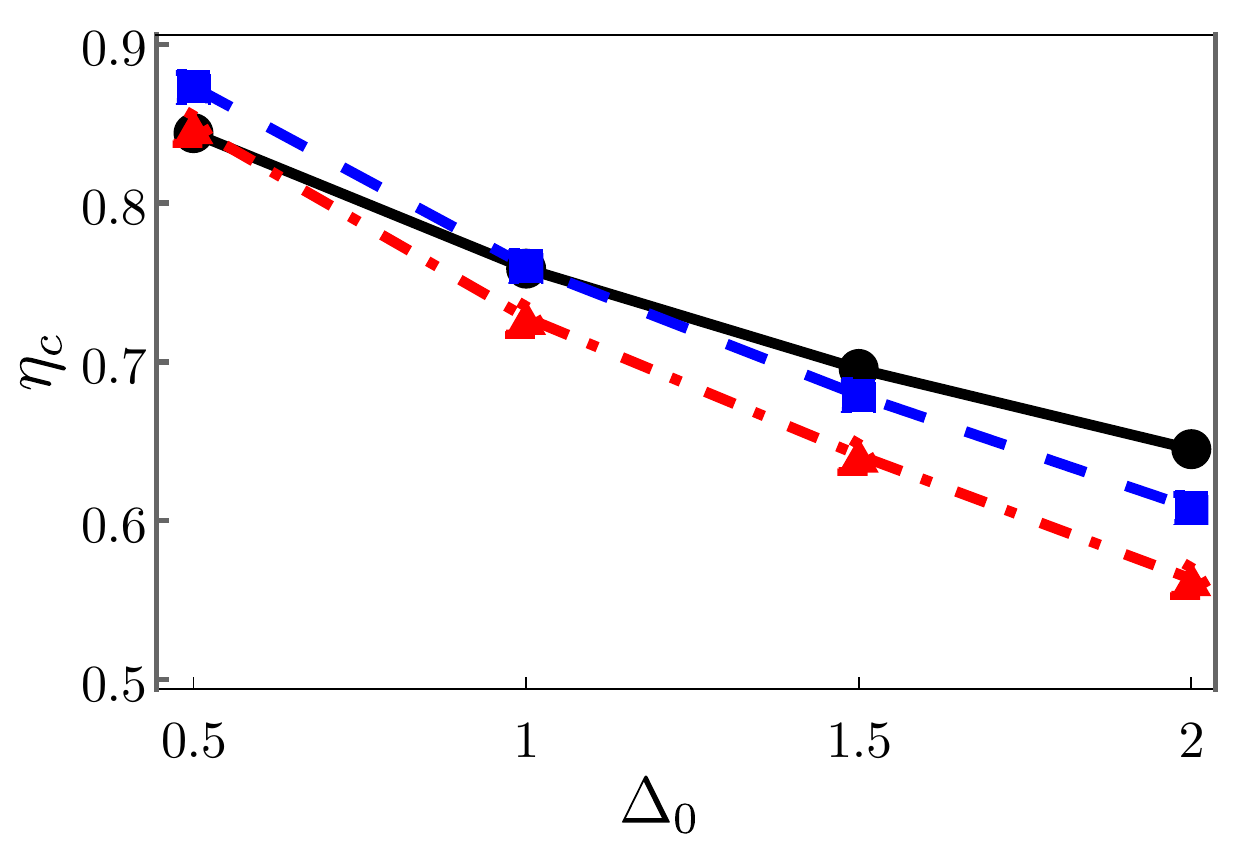}
\end{center}
\vspace*{-4mm}
\caption{
Efficiency $\eta_{c}$ at $\alpha_{c}(g,\Delta_0)$ versus $\Delta_0$. Nonlinearities:
attractive $g=-1$ (black solid), non-interaction $g=0$ (blue dashed), and repulsive $g=1$ (dot-dashed red).
Values of $\alpha_{c}(g,\Delta_0)$ are the same as those in Fig. \ref{fig3:alphac_delta}.
\label{fig5:effi_delta}}
\end{figure}

Figure \ref{fig4:effi_gs_pert} shows that the efficiency $\eta$ grows as $\alpha$ increases. The approximate efficiency in Eq. (\ref{eff_pert}) is a quadratic function of $\alpha$, and this is in good agreement with the numerical results in Fig. \ref{fig4:effi_gs_pert}(a) for the case $g=0$.
In the limit of $\Delta_0 \rightarrow 0$, the efficiency $\eta \sim \alpha^{2}$,
see 
Eq. (\ref{zeroDelta}). This limit case is also shown by the dashed pink line in Fig. \ref{fig4:effi_gs_pert}(a).
As one expects a maximum of the total work close to  $\alpha_{c}(g,\Delta_0)$, one expects also that the efficiency reaches the maximum at $\alpha$ close to 
$\alpha_{c}(g,\Delta_0)$.
The efficiency $\eta_c$ at a critical $\alpha_c$ with respect to $\Delta_0$ is shown in Fig. \ref{fig5:effi_delta}. 
The efficiency decreases with increasing $\Delta$. This corresponds to Eq. (\ref{eff_pert}) when $\alpha=\alpha_c \approx \pi/a_2$ for all three cases of $g$ (see Fig. (\ref{fig3:alphac_delta})).

\textit{Discussion and conclusions} Here we return to the physical units and discuss the possibility of experimental realization of the present Stirling cycle. 
In the one-dimensional realization considered above, with the physical unit of length $\xi$, the resulting dimensionless coupling constant $g$ can be estimated as  $\sim\,2{\cal N}a_{\rm at}\xi/s_{\rm p},$ where $s_{\rm p}$ is the condensate cross-section, physically corresponding to the piston cross-section. 
Here $a_{\rm at}$ is the interatomic scattering length  
(typically of the order of $10a_{B}$, where $a_{B}$ is the Bohr radius) dependent on the Feshbach resonance realization, and ${\cal N}\sim 10^{3}$ is the total number of atoms in the condensate. 
A reasonable $\xi$ for optical setups is of the order of 10 $\mu$m. Thus, the choice of $a_{1}, a_{2}$ of the order of 10 $\mu$m allows one to achieve dimensionless $\alpha$ and $\Delta_{0}$ of the order of unity \cite{PhysRevA.84.025602}, and thus explore the operational regimes of the Stirling cycle up to the critical values.

In summary, we have explored the potential of a spin-orbit coupled Bose-Einstein condensate 
in a thermodynamic Stirling-like cycle. 
It takes advantage of both the non-commuting synthetic spin-orbit and Zeeman-like contributions.
The "cooling" and "heating" is assumed to originate by interaction with external magnetization and demagnetization media.
We have examined the ground-state energy of the condensate and how the corresponding pressure depends on the different parameters of the system.
We have studied the efficiency of the corresponding engine in the dependence on the strength of these spin-related couplings. 
The cycle is characterized by a critical spin-orbit coupling, corresponding, essentially, to the maximum efficiency. 
The dependence of the efficiency on the spin-dependent coupling and nonlinear self-interaction paves the way to applications of these cycles. 
While we have concentrated here on effects originating from the BEC, it will be
interesting to study the details of the effects of the external magnetization and demagnetization sources in the future.

\begin{acknowledgements}
We are grateful to C. Whitty and D. Rea for commenting on the manuscript. 
J.L. and A.R. acknowledge support from the Science Foundation Ireland Frontiers for the Future Research Grant ``Shortcut-Enhanced Quantum Thermodynamics" No.19/FFP/6951. 
The work of E.S. is financially supported through the Grant PGC2018-101355-B-I00 funded by MCIN/AEI/10.13039/501100011033 
and by ERDF ``A way of making Europe'', and by the Basque Government through Grant No. IT986-16.
\end{acknowledgements}

%



\end{document}